\documentclass[sigconf]{acmart}

\AtBeginDocument{%
  \providecommand\BibTeX{{%
    \normalfont B\kern-0.5em{\scshape i\kern-0.25em b}\kern-0.8em\TeX}}}

\usepackage{amsmath}
\usepackage{svg}
\usepackage{graphicx}
\usepackage{caption}
\usepackage{subcaption}
\usepackage{algorithm}
\usepackage[noend]{algpseudocode}

\newcommand{\specialcell}[2][c]{%
\begin{tabular}[#1]{@{}c@{}}#2\end{tabular}}

\begin{document}

\title{ORB-based SLAM accelerator on SoC FPGA}

\author{Vibhakar Vemulapati}
\affiliation{%
  \institution{University of Illinois at Urbana-Champaign}
  \city{Urbana}
  \state{IL}
  \postcode{61801}
}\email{vemulpt2@illinois.edu}

\author{Deming Chen}
\affiliation{%
  \institution{University of Illinois at Urbana-Champaign}
  \city{Urbana}
  \state{IL}
  \postcode{61801}
}\email{dchen@illinois.edu}

\renewcommand{\shortauthors}{Vemulapati, et al.}

\begin{abstract}
Simultaneous Localization and Mapping (SLAM) is one of the main components of autonomous navigation systems. With the increase in popularity of drones, autonomous navigation on low-power systems is seeing widespread application. Most SLAM algorithms are computationally intensive and struggle to run in real-time on embedded devices with reasonable accuracy. ORB-SLAM is an open-sourced feature-based SLAM that achieves high accuracy with reduced computational complexity. We propose an SoC based ORB-SLAM system that accelerates the computationally intensive visual feature extraction and matching on hardware. Our FPGA system based on a Zynq-family SoC runs 8.5x, 1.55x and 1.35x faster compared to an ARM CPU, Intel Desktop CPU, and a state-of-the-art FPGA system respectively, while averaging a 2x improvement in accuracy compared to prior work on FPGA.
\end{abstract}


\maketitle

    \section{Introduction}
    \label{section:introduction}
    Simultaneous Localization and Mapping (SLAM) is the problem of constructing a map of an unknown environment, while simultaneously estimating the current pose (position and orientation w.r.t a fixed reference) of the observer. The mapping and pose need to be solved simultaneously because pose estimates from odometry alone are subject to noise and drift, while constructing a map requires a good estimate of your current pose. SLAM is a fundamental problem in autonomous navigation and location estimation with applications in autonomous cars/drones and augmented reality. SLAM has the potential opportunity of replacing or enhancing GPS tracking and navigation in certain applications and environments. GPS systems are not accurate indoors or in big cities, where there are obstructions between the navigation satellites and the observer. Moreover, in the best of conditions, GPS is only accurate to within a meter in the horizontal plane and around 3 meters in the vertical direction. SLAM offers the benefit of not relying on external beacons (satellites) in unreliable environments while potentially being more accurate about the observer's location.
    
    Visual SLAM solves the SLAM problem using visual data for both pose estimation and mapping. The visual data can be either from monocular images, stereo images, or 3-D images. With the rise in popularity in UAVs (unmanned aerial vehicles), visual SLAM algorithms are gaining traction in embedded systems that run in low-power environments, especially if they can perform well with low-quality sensors. Visual SLAM algorithms can be broadly categorized based on two criteria: the density of the map being reconstructed and the type of inputs being tracked. Reconstructed maps can be dense, which reconstruct the entire space, semi-dense, which reconstruct detected edges, and sparse, which only reconstruct features. Features are points of interest (generally corners) in each frame that are tracked across frames as points of reference. Visual SLAM algorithms either attempt to track every pixel in the frame or attempt to track features detected.

    One of such visual SLAM algorithms published as an open source library is ORB-SLAM \cite{mur2015orb}\cite{mur2017orb}\cite{ORBSLAM3_TRO}. ORB-SLAM utilizes ORB (Oriented FAST, Rotated Brief) features explained in more detail in Section \ref{section:ORB}. It gained significant popularity in the visual SLAM community for implementing a highly accurate monocular SLAM system. Later, the work was extended to utilize stereo images \cite{mur2017orb}, as well as RGB-D images, which at the time was one of the most accurate visual SLAM algorithms publicly available. ORB-SLAM proposes a feature-based sparse visual SLAM, making it relatively less computationally intensive (compared to other SLAM systems). Inspite of the that, ORB-SLAM still struggles to run on embedded devices. Table \ref{tab:orbslam_performance} shows the performance of ORB-SLAM on various platforms when using RGB-D inputs. NVIDIA's Jetson Xavier platform, which sports a large 8-core ARMv8.2 and is marketed towards robotics applications, is only able to achieve 10 frames per second, using purely the CPU, and 22fps when utilizing the CPU and the GPU \cite{aldegheri2020}. We also observe that the feature extraction component accounts for 60-70\% of the runtime of the algorithm. Feature extraction is an image processing algorithm, which involves many parallel computation patterns, making it well suited for FPGAs. 
    
    \begin{table}[h]
        \caption{Average Performance of ORB SLAM on various platforms. "ORB runtime" indicates the time of the frame spent extracting ORB features.}
        \label{tab:orbslam_performance}
        \centering
        \begin{tabular}{|c|c|c|c|}
            \toprule
            Device & FPS & \specialcell{Total Frame \\runtime (ms)} & \specialcell{ORB runtime \\(ms)}\\
            \midrule
             Intel i5 Desktop & 40 & 25 & 16\\
             Intel i5 Laptop & 33 & 30 & 18\\
             NVIDIA Jetson (CPU) & 10 & 100 & 70 \\
             NVIDIA Jetson (CPU+GPU) & 22 & 45 & 20 \\
             ARM CPU & 7 & 142 & 100\\
             \bottomrule
        \end{tabular}
    \end{table}
    
    SLAM acceleration using FPGAs is a difficult problem as there are multiple constraints that need to be met. The system needs to have real-time performance, high accuracy, and be able to fit within an embedded device. There has been some work in accelerating ORB feature extraction by itself, which is evaluated by accuracy of tracking features across frames. Previous works have experimented with hardware optimizations of ORB feature extraction and matching by trading off accuracy for performance. The accuracy of SLAM systems is inherently noisy, and hence, the performance of ORB feature matching does not necessarily correlate with the end performance of visual SLAM. In this work, we propose hardware optimizations to improve the run-time performance on embedded devices. We investigate how these hardware optimizations affect the final accuracy of the SLAM system, ignoring the accuracy of intermediate computations, allowing us to make global optimizations that have been overlooked.
    The contributions of this work can be summarized as follows:
    \begin{itemize}
        \item Implementation of an end-to-end SLAM accelerator that runs on an FPGA SoC.
        \item A hardware-based accelerator for the ORB feature extraction, which can support multiple image resolutions and multiple levels of image pyramid processing in parallel.
        \item Data-driven optimizations and approximations such as bit-width pruning and data quantization to ensure resource efficiency, while minimizing impact on the accuracy of the overall system.
    \end{itemize}
    
    The rest of the paper is organized as follows: Section \ref{section:related_work} provides an overview of prior implementations of ORB and ORB-SLAM algorithms in hardware. Section \ref{section:ORB-SLAM} introduces the ORB-SLAM algorithm and briefly details the various components and explains the ORB algorithm in detail. Section \ref{section:hardware_architecture} describes the hardware architecture of our ORB accelerator. Section \ref{section:experiments} highlights the experimental results that guided our optimizations. Section \ref{section:conclusion} summarizes our results and insights in this paper.

    \section{Related Work}
    \label{section:related_work}
    FPGAs have become popular in the embedded device space as they enable real-time performance while maintaining a low energy profile compared to embedded CPUs or GPUs. Applications that can be broken down into parallelizable tasks such as vision-based object detection, sound localization, machine learning inference and image classification have shown great results on FPGAs compared to other systems. Liu, et al. \cite{liu2011real} proposed an FPGA accelerator for real-time object detection and Zhao, et al. \cite{zhao2012real} proposed a GPU accelerator for 3-D sound localization. Both works exploit the inherent parallelism in the problems to achieve large speedups when compared to their CPU counterparts. 
    
    With the proliferation of Internet of Things (IoT) devices in our day-to-day lives, the industry is seeing more applications that require low power and high performance. Chen, et al. \cite{chen2016platform} detailed on the various challenges that the industry faces in choosing the appropriate platform for IoT devices while also proposing design flows and software tools to help address these challenges. The paper makes a strong case for SoC-FPGAs to be utilized when the application requirements are aggressive in performance and power.

    The major drawbacks in deploying FPGA accelerators compared to CPUs or GPUs are constrained resources and longer development times. FPGAs have far fewer arithmetic units, memories and logical elements when compared to GPUs. A general strategy for migrating applications to FPGAs involves optimizing the algorithm to fit on resource-constrained devices while maintaining real-time performance. Zhang et al. \cite{zhang2019skynet} and Li, et al. \cite{li2020edd} propose modifications such as bit-width pruning, data quantization and software-hardware co-design to fit the applications on embedded devices without sacrificing much accuracy. 
    
    FPGA accelerator development typically needs specialized knowledge in both FPGA design as well as the targeted application making it expensive and time-consuming to develop and verify designs. Reducing the development time and cost to deploy FPGA accelerators would require major enhancements to current CAD software toolchains. Hao, et al. \cite{hao2018deep}\cite{hao2019fpga}, Xu, et al. \cite{xu2020autodnnchip} and Zhang, et al. \cite{zhang2018dnnbuilder} proposed automated flows to generate hardware designs given high-level specifications of deep neural networks. They utilize various design-space exploration techniques to simultaneously optimize the accuracy, performance and resource utilization of the application to automatically generate the best implementation for edge computing devices without the need for manual design effort.

    SLAM algorithms are parallelizable and energy intensive, and hence are well suited to FPGA acceleration. There have been several prior works that have investigated SLAM acceleration on FPGAs. The works can be broadly categorized into two buckets: The front-end ORB feature extraction accelerators, and back-end Bundle Adjustment accelerators. 
    Fang, et al. \cite{fang2017fpga} implemented an ORB accelerator to integrate into their visual SLAM flow. They were able to achieve 67 fps on 640x480 images on an FPGA running at 200MHz. They optimized some resources by truncating the width of the intermediate values of computation.  
    Weberruss et al. \cite{weberruss2017fpga} proposed a multi-scale ORB accelerator. They achieved 72fps on 1920x1080 images, processing the image at 1 pixel/cycle.
    Tran et al. \cite{tran2018stream} propose a stream-based ORB accelerator focusing on dynamic power optimizations using dynamic clock gating, and threshold-guided bit-width pruning of intermediate computation values. No performance numbers were reported. Both referenced studies \cite{tran2018stream, weberruss2017fpga} use the Harris Corner Detection algorithm instead of FAST for corner detection. Although Harris Corner is more complex computationally, it provides better accuracy in corner extraction.
    Qin, et al. \cite{qin2019pi} propose an FPGA SoC based accelerator to target the Bundle adjustment portion of SLAM. They proposed a hardware/software co-designed accelerator which sped up the Schur Complement computation of the algorithm. Their work achieved a speedup of 1.5x compared to the ARM implementation. The authors further extended on this work in \cite{liu2020pi} and improved the speedup to 7.5x compared to ARM.
    Wu, et al. \cite{wu2021fpga} propose an FPGA accelerator for DS-SLAM, a neural network based SLAM \cite{yu2018ds}. Neural-network based SLAMs have been emerging in the field, as they are more accurate; however, the computational cost is an order of magnitude higher. The authors implement a platform called HERO that is 5x more energy efficient than a desktop system, while running at 5fps.
    Liu, et al. \cite{liu2021archytas} propose a framework for generating hardware for localization acceleration. They are able to achieve speed-ups upto 8x over Desktop CPUs, while consuming a great amount of resources.
    
    A few works implement end-to-end SLAM accelerators. Boikos, et al. \cite{boikos2017high} implement the full LSD-SLAM pipeline on a Zynq SOC. LSD-SLAM is a semi-dense direct visual SLAM algorithm \cite{engel2014lsd}. They offload compute-intensive optimization problems onto the FPGA while the CPU takes care of bookkeeping and other minor tasks. They were able to achieve 22fps on 640x480 images. Asgari, et al. \cite{asgari2020pisces} proposed an end-to-end ORB-based SLAM system that uses EKF (Extended Kalman filters) as the optimization backend. EKFs have not been used in the robotics community for a few years, as graph-based approaches have taken over the back-end of SLAM. The authors implement a system that is more power efficient and faster than eSLAM \cite{liu2019eslam}, but fail to mention accuracy in the paper, so it is difficult to gauge if it performs in real-world scenarios. Liu, et al. \cite{liu2019eslam} implement a visual SLAM system based on ORB features on a Zynq SoC. Feature extraction and matching was offloaded to the FPGA, while the pose optimization, pose estimation and map updating were performed on the CPU. They also proposed a hardware-friendly optimization to BRIEF patterns, which costs fewer FPGA resources, but produces a less accurate system.
    
    We compare the hardware implementations of the above works in more detail in Section \ref{section:other_works}.

    \section{ORB-SLAM}
    \label{section:ORB-SLAM}
    \subsection{SLAM}
     SLAM algorithms consist of two parts: map generation and localization. The map generation places points in 3-d space, which can then be used as reference points for localization. Localization estimates the pose of the camera by observing the changes in features across consecutive frames. Modern visual SLAM algorithms use a keyframe based approach where the keyframes contain information about the map points they observe. A new keyframe is generated when either a certain amount of time has passed or a certain number of new points are observed that have not been observed in other keyframes. Localization is performed using the position estimates of features observed in the current frame that are observed in adjacent keyframes using a bundle adjustment algorithm detailed further in Section \ref{section:bundle_adjustment}. Keyframe based approaches use a graph-based method for map point and pose estimation. In contrast, earlier SLAM algorithms used a filter-based approach, where localization and mapping were performed simultaneously. Graph-based SLAM has become more popular recently with works like Grisetti, et al. \cite{grisetti2010tutorial} explaining why it is more accurate.
    
    ORB-SLAM is a sparse feature-based visual SLAM algorithm capable of operating with monocular, stereo, and monocular + depth inputs. The algorithm consists of three main threads that are running in parallel: 
    \begin{itemize}
        \item Tracking: Extracts ORB features, performs pose estimation and adds keyframes.
        \item Local Mapping: Updates the map with new points from keyframe. Performs optimization of local keyframes (Local Bundle Adjustment). Removes redundant keyframes and bad map points.
        \item Loop Closing: Checks to see if an area has been revisited, allowing the algorithm to re-calibrate the pose and correct any drift accumulated over time. Optimizes all the keyframes (Global Bundle Adjustment) on loop closure. It is an optional feature that greatly enhances accuracy in scenarios where the same location is visited multiple times.
    \end{itemize}
    
    \subsection{Bundle adjustment (BA)} 
    \label{section:bundle_adjustment}
    
    \begin{figure}[h]
      \centering
      \includegraphics[width=0.8\linewidth]{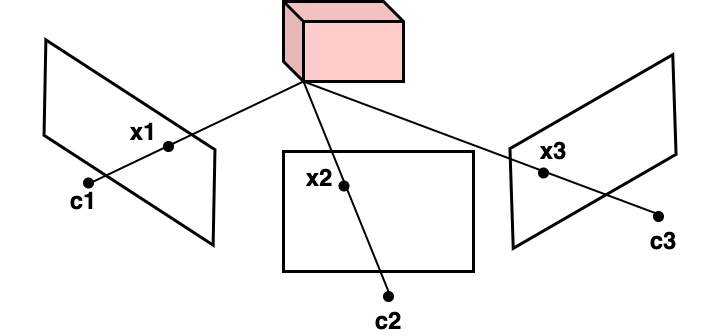}
      \caption{Bundle adjustment estimation example. The same point on the cuboid is observed from 3 different camera positions $c1$, $c2$, $c3$. The corner is projected onto the 2-d frame at $x1$, $x2$, $x3$.}
      \label{fig:bundle_adjustment}
    \end{figure}
    
    Bundle adjustment(BA) is an algorithm in photogrammetry used to reconstruct an image from multiple viewpoints \cite{triggs1999bundle}. A 3-D point is reprojected from the camera's 2-D reference frame into the world coordinates based on multiple observations from different viewpoints. As in Figure \ref{fig:bundle_adjustment}, the same point on the cuboid is observed from multiple frames, with each observation having inherent uncertainty. Since there are more equations than variables, BA attempts to minimize the reprojection error, across all equations using a non-linear regression. ORB-SLAM has 3 types of BA, each running on independant threads: Motion-only BA, local BA, and global BA.
    \begin{itemize}
        \item Motion only BA optimizes the camera's position and orientation with respect to the features observed in the current frame. It does not update the map points. For given position $t \in (x,y,z)$ and orientation $ R \in SO(3)$ (3-D rotation group), we optimize for pose using the following equation, which aims to minimize the error between observed and predicted position of every point in the frame.
        $$  \{R,t\} = argmin\sum_{i\in \chi} ||x^i - X^i(R,t)||^2 $$
         $\chi$ is the set of all features of the current frame matched with the map points, with $x$ being a matched feature, and $X$, the projection of the matched map point into the current frame.
        \item Local BA optimizes all the map points observed in a given set of keyframes that share observations. It is executed on the addition of a new keyframe.
        \item Global BA optimizes all the keyframes detected so far (with the exception of the first keyframe). This is done when a loop is detected to correct the observations of all the map points.
    \end{itemize}

    \subsection{ORB}
    \label{section:ORB}
\begin{figure}[h]
      \centering
      \includegraphics[width=\linewidth]{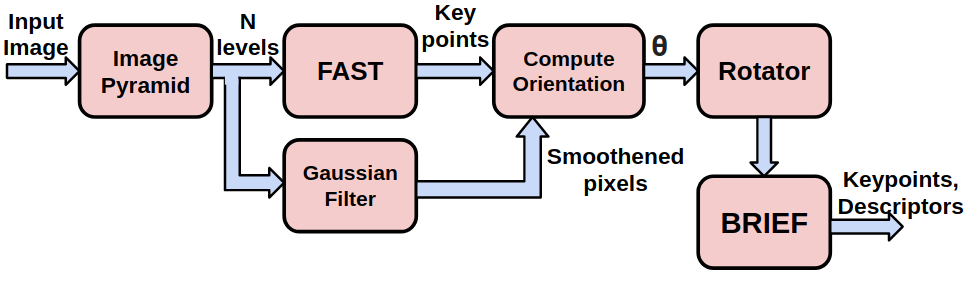}
      \caption{ORB Pipeline}
      \label{fig:orb_pipeline}
    \end{figure}

    ORB \cite {rublee2011orb} is a feature extraction and description algorithm that efficiently identifies features in an image and generates a unique descriptor for each feature in order to identify them in other frames. ORB has been proven to be robust, while maintaining a relatively simple computational pipeline, making it attractive for feature-based sparse SLAM algorithms. ORB (Oriented-FAST and Rotated-BRIEF) can be broken down into two parts: feature extraction (oriented-FAST) and feature description (rotated-BRIEF). oFAST extracts corners and determines their orientation. rBRIEF generates a rotation invariant descriptor. The ORB computation pipeline is shown in Figure \ref{fig:orb_pipeline}. The algorithm takes in the image, and outputs the keypoints and their associated descriptors. In the following sections, the ORB algorithm will be described in more detail.
    
    \subsection{Image Pyramid}
    ORB extracts features from multiple scales of an image: the original image, and multiple downsampled versions of the original. Feature extraction algorithms are usually implemented with a fixed-size sliding window over the input image. Multiple image scales effectively increase the size of the sliding window, allowing it to detect features, that would otherwise, not be seen in different scales. For instance, large, rounded corners would be missed in larger images, while at smaller scales, they get shrunk down to a detectable size. Image pyramids also enhance the efficiency when the input camera is in motion, where the same object would be detected at multiple scales as the camera moves towards/away from it.
     The image pyramid used in the ORB-SLAM pipeline consists of 4 levels, with each level 1.2 times smaller than the previous.
    \subsection{FAST}
    
\begin{figure}[h]
        
        \centering
         \begin{subfigure}[b]{0.3\linewidth}
             \centering
             \includegraphics[width=\linewidth]{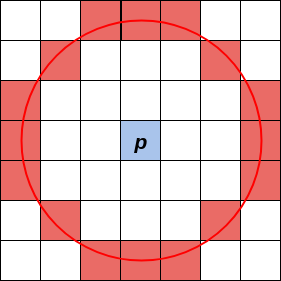}
             \subcaption{Bresenham circle of radius 3 around pixel $p$}
             \label{fig:bresenham}
         \end{subfigure}
         \hfill
         \begin{subfigure}[b]{0.6\linewidth}
             \centering
             \includegraphics[width=\linewidth]{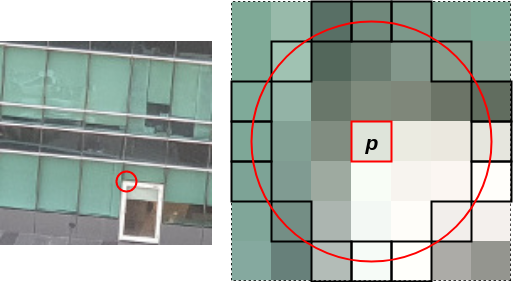}
             \subcaption{Example of an image with the surrounding Bresenham circle}
             \label{fig:fast_example}
         \end{subfigure}
         \hfill
                  \caption{Bresenham circle used in FAST corner detection}

     \end{figure}

    FAST examines the local environment of a pixel $p$ to determine whether it is a corner. It examines a ring of 16 pixels, located on a Bresenham circle of radius 3, around $p$ as shown in Figure \ref{fig:bresenham} and \ref{fig:fast_example}. 
    A pixel is said to be a corner if 9 contiguous pixels in the circle are either all brighter than, or all darker than $p$ above/below the given threshold. 
    
    The FAST algorithm is followed by a non-maximum suppression (NMS) of neighbouring pixels to avoid duplication of corners that are in proximity. A score is computed for each pixel $p$ as the sum of the absolute differences between $p$ and the pixels in the Bresenham circle and passed on to the NMS module. The corner with the highest score in the window is determined to be a keypoint, while the others are suppressed.
    
    \subsection{Orientation}
    The orientation of a keypoint is determined using an intensity centroid of the surronding pixels. First, the image is smoothened using a Gaussian filter of size 7x7. The centroid of image patch can be calculated by defining the moments in the x and y ($m_{01}$ and $m_{10}$ respectively) directions of the image patch. The moments give us the coordinates of the centroid, which can be used to calculate the orientation $\theta$ of the keypoint.
    
    \begin{equation}\label{eq:moment} m_{pq} = \sum_{x,y} x^p y^q I(x,y) \end{equation}
    
    $$ C = \left ( \frac{m_{10}}{m_{00}}, \frac{m_{01}}{m_{00}}\right )$$
    
    $$ \theta = atan2( \frac{m_{01}}{m_{00}}, \frac{m_{10}}{m_{00}}) = atan2(m_{01}, m_{01}) $$
    
    The ORB implementation \cite{rublee2011orb} suggests that a 31x31 image patch around the keypoint is suitable to compute a robust orientation. 
    
    \subsection {BRIEF}
    \label{section:BRIEF}
\begin{figure}[h]
         \centering
         \begin{subfigure}[b]{0.3\linewidth}
             \centering
             \includegraphics[width=\linewidth]{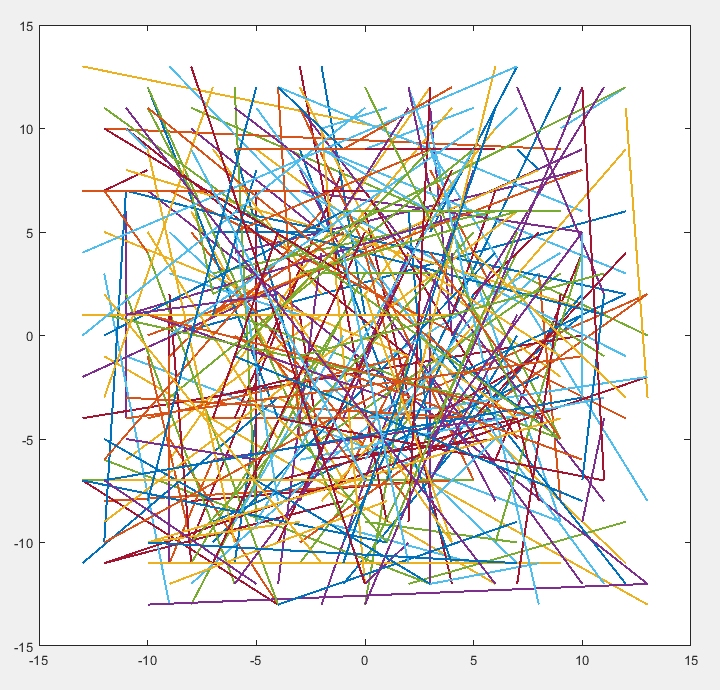}
             \caption{}
             \label{fig:brief_random_3}
         \end{subfigure}
         \hfill
         \begin{subfigure}[b]{0.3\linewidth}
             \centering
             \includegraphics[width=\linewidth]{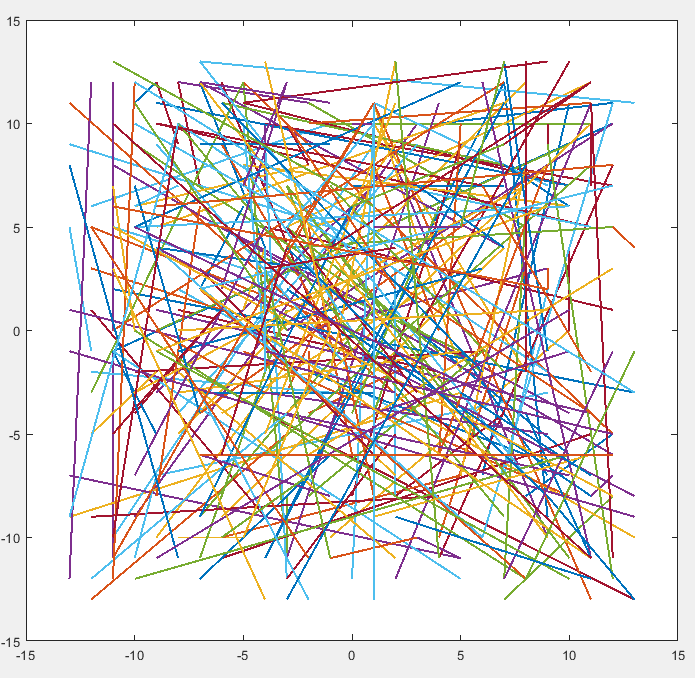}
             \caption{}
             \label{fig:brief_random_2}
         \end{subfigure}
         \hfill
         \begin{subfigure}[b]{0.3\linewidth}
             \centering
             \includegraphics[width=\linewidth]{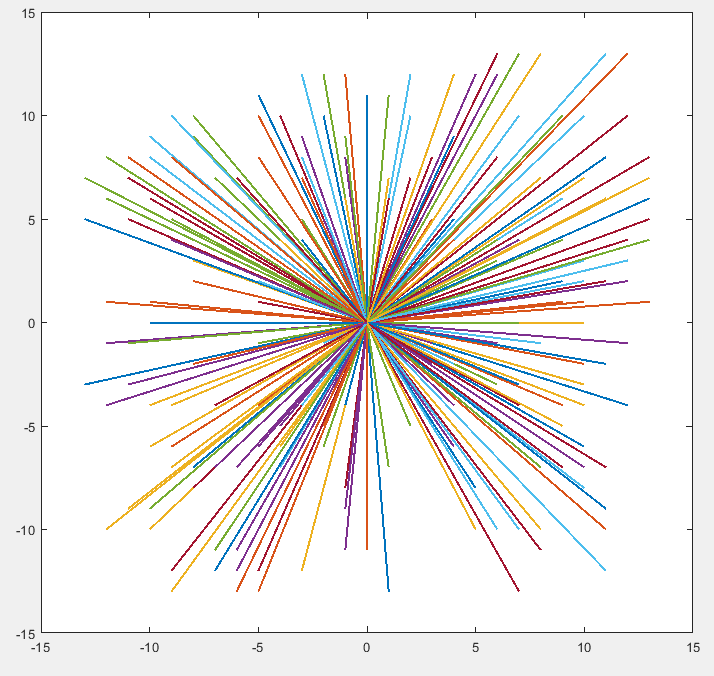}
             \caption{}
             \label{fig:brief_polar}
         \end{subfigure}
            \caption{Example BRIEF patterns. Each line connects a pair of points that are compared to determine the descriptor. Axes represent coordinates relative to center point}
            \label{fig:brief_pattern}
    \end{figure}

    BRIEF \cite{calonder2010brief} outputs a description of each keypoint, as a unique identifier, for use in other frames (like a hash). The algorithm performs $N$ comparisons between pairs of pixels and outputs an $N$-bit binary string as the output. The original ORB algorithm suggests choosing the pixel pairs as highly random with low correlation amongst the pairs, to maximize success and minimize false positives while matching keypoints. Figure \ref{fig:brief_pattern} illustrates various BRIEF patterns. Figure \ref{fig:brief_polar}  exhibits very low randomness in pair selection, which will reduce the uniqueness of the descriptors, while Figure \ref{fig:brief_random_2} and \ref{fig:brief_random_3} are more likely to generate a unique descriptor.
    
    The descriptor is calculated by selecting $N$ pixel pairs (A,B) and setting the output bit if the intensity of pixel A is greater than pixel B, and reset otherwise.
    
    $$\text{descriptor[}i\text{]} = I(A) > I(B)$$
    where $i = 0 ... N$, $A$ and $B$ are sets of coordinates.
    
    To achieve rotation invariance, the orientation angle of the keypoint, $\theta$, is used to rotate the each pair in the BRIEF pattern using a rotation matrix.
    
    \begin{equation}
    \label{eq:brief_rotation}
    \begin {bmatrix}
        x' \\
        y'
        \end {bmatrix} = 
        \begin{bmatrix} 
    cos\theta & -sin\theta \\
    sin\theta & cos\theta
    \end{bmatrix}
        \begin {bmatrix} 
        x\\
        y
        \end{bmatrix}
    \end{equation}

    The pixels are then sampled using the new coordinates $x'$ and $y'$, to generate the descriptor. As suggested in the BRIEF paper \cite{calonder2010brief}, a 31x31 patch surrounding the keypoint is sufficient for uniqueness. However, since we have to rotate the BRIEF patterns, once rotated, the coordinates can now lie outside the bounds of the patch. To accommodate for this, we limit our BRIEF pattern to a 27x27 patch, and, once it is rotated, all the points will lie within a 37x37 patch. ORB uses a 256-bit descriptor, as it was determined to generate a unique enough descriptor while maintaining a low length.
    
    If a feature is detected in multiple frames, from different view points or orientations, their BRIEF patterns would be similar. Nearest neighbor search using Hamming distance is used to match the same feature across frames.

    \section {Hardware architecture}
    \label{section:hardware_architecture}
\begin{figure}[h]
        \centering
        \includegraphics[width=0.4\linewidth]{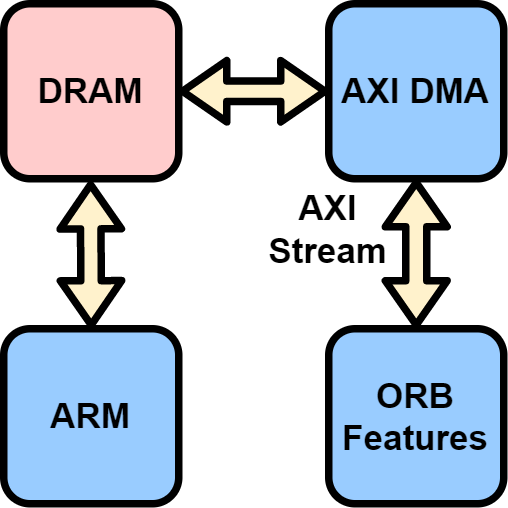}
        \caption{Top level architecture of the ORB-SLAM implementation. Computation is partitioned between software and hardware.}
        \label{fig:SOC_toplevel}
    \end{figure}

    The architecture of the proposed accelerator is shown in Figure \ref{fig:SOC_toplevel}. ORB feature extraction and matching consumes 60-70\% of the runtime of the algorithm, and, hence, it is offloaded onto hardware. The rest of the algorithm (Bundle adjustment) is run in software on the ARM cores. Based on the data from previous works, accelerating both the Bundle Adjustment and the ORB on the same FPGA would require a tremendous amount of resources. 
    
    \begin{figure}[h]
        \centering
        \includegraphics[width=\linewidth]{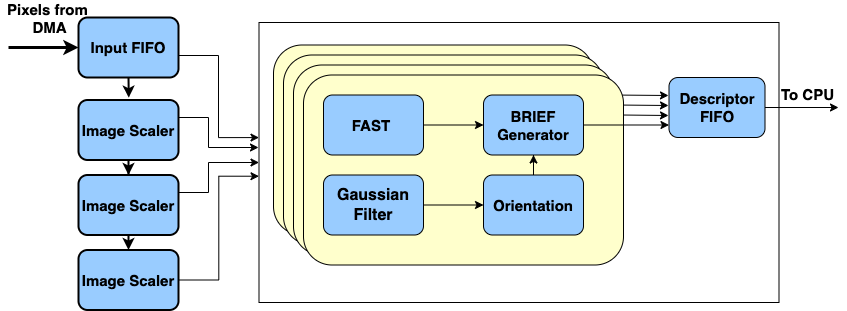}
        \caption{ORB Feature extraction module}
        \label{fig:FPGA_toplevel}
    \end{figure}
    
    Figure \ref{fig:FPGA_toplevel} illustrates the top-level architecture of the ORB feature extraction accelerator.
    The software and hardware are coupled using AXI-Stream via an AXI DMA. The feature extraction module operates as a streaming processor to take advantage of data re-use, without requiring the use of DRAM to store intermediate state. The image cannot be stored in a frame buffer, as an entire image will not fit in most embedded FPGA's BRAM. The input image pixels are streamed from DRAM to the FPGA into a FIFO which are then dispatched to the ORB extractor. After the features are extracted, the descriptors are stored in a FIFO until transferred back to the CPU. The modules in hardware are fully pipelined, allowing it to process 1 pixel/cycle when the inputs are saturated. The feature extraction and matching can be completely independent of the back-end localization. This allows us to pipeline the dataflow between the front-end and back-end of the system. In the following sections, we will describe the hardware architecture of the ORB feature extraction module on the Programmable Logic.
    
    \subsection {Buffers}
\begin{figure}[h]
        \centering
        \includegraphics[width=\linewidth]{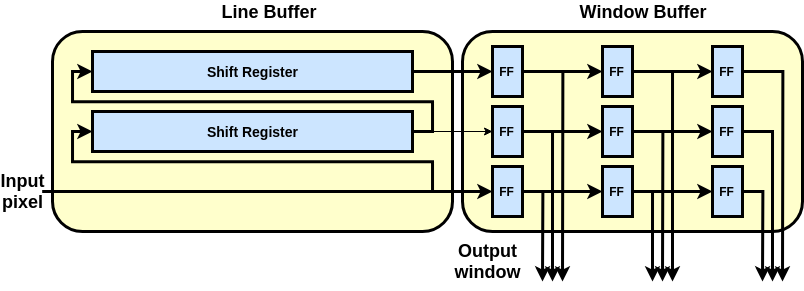}
        \caption{Linebuffer of 3 rows followed by a 3x3 window}
        \label{fig:line_buffers}
    \end{figure}

    Throughout the pipeline we make use of buffers in two forms: line buffers, and window buffers. Pixels are streamed in a row major fashion; however, the kernels require input from multiple rows simultaneously to calculate the output for one pixel. We use line buffers as a mechanism to delay the input by N rows so that we have access to all N rows simultaneously. Figure \ref{fig:line_buffers} illustrates an example of the line buffers, which can be built from either BRAMs or the Shift Register primitives provided by LUTs. Window buffers are utilized to access pixel data in parallel. It is set up as 2-dimensional array of shift registers, and generally follows a linebuffer.
    
    \subsection {Image scalers}
    To generate our image pyramid, we downsample the input image at multiple levels. In order to avoid jagged edges and loss of information, the image has to be resampled. ORB-SLAM uses a scale factor of 1.2 for downsampling.  This can be expressed as a fraction $\frac{5}{6}$.
    Bilinear interpolation is used for resampled pixels that lie between the original pixels. The interpolation can be implemented using the 4 neighboring pixels with weights based on the location of the resampled pixel. The resampled pixel can be in one of 25 different locations, giving us 25 sets of weights. The weights can be calculated purely as a function of the original image's x and y coordinate modulo 6.
    
    $$
        \begin{bmatrix}
            x_6 * y_6 & (5 - x_6) * y_6\\
            x_6 * (5 - y_6) & (5 - x_6) * (5 - y_6) \\
        \end{bmatrix}
    $$
    where $x_6 = x \text{ modulo } 6$,  $y_6 = y \text{ modulo } 6$. This makes an efficient down scaler only require 1 linebuffer and a 2x2 window buffer. If either $x_6$ or $y_6$ is equal to 5, the resampled pixel is not valid.  Note that the downscaled pixels are only outputted every 5 out of 6 cycles, and no pixels are outputted on every 6th row.
    

    \subsection {FAST}
    The FAST corner detection algorithm is very suitable for FPGAs, as all the comparisons required can take place in parallel. We use a multi-stage pipeline to compute the pixel differences, compare the threshold, and generate two 16-bit vectors representing whether each pixel in the Bresenham circle is either darker or brighter than pixel $p$ as seen in Figure \ref{fig:bresenham}. Sixteen 16-bit bitmasks are generated, for each combination of 9 contiguous ones in the vector. The dark and bright vectors are ANDed with the bitmasks, and tested for equality. If the pixel is either brighter than or darker than its neighbours it is declared as corner. At the same time, the pixel score for $p$ is computed. If a $p$ is not a corner, FAST outputs 0; otherwise, it outputs the pixel score.
    
    \begin{figure}[h]
     \centering
     \includegraphics[width=0.8\linewidth]{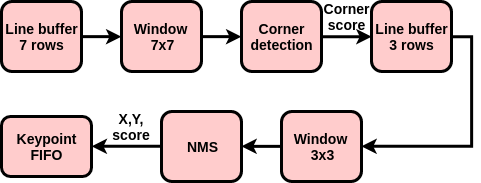}
     \caption{Top level block diagram of FAST}
     \label{fig:fast_block_diagram}
    \end{figure}

    
    After the FAST detection, the pixel scores are stored in a linebuffer with two rows as shown in Figure \ref{fig:fast_block_diagram}.
    NMS (non-max suppression) then compares the pixel $p$ with all the other pixels in a 3x3 window. If $p$'s score is greater than that of all the neighboring pixels, it is considered a keypoint. When comparing against pixels to the right and down of $p$, NMS checks if $p$ is greater than or equal to those neighbors, to avoid suppression of adjacent corners that have the same corner score. The keypoints are written into the keypoint FIFO along with its (x,y) coordinate.

    \subsection {Gaussian Filter}
\begin{figure}[h]
     \centering
     \includegraphics[width=0.65\linewidth]{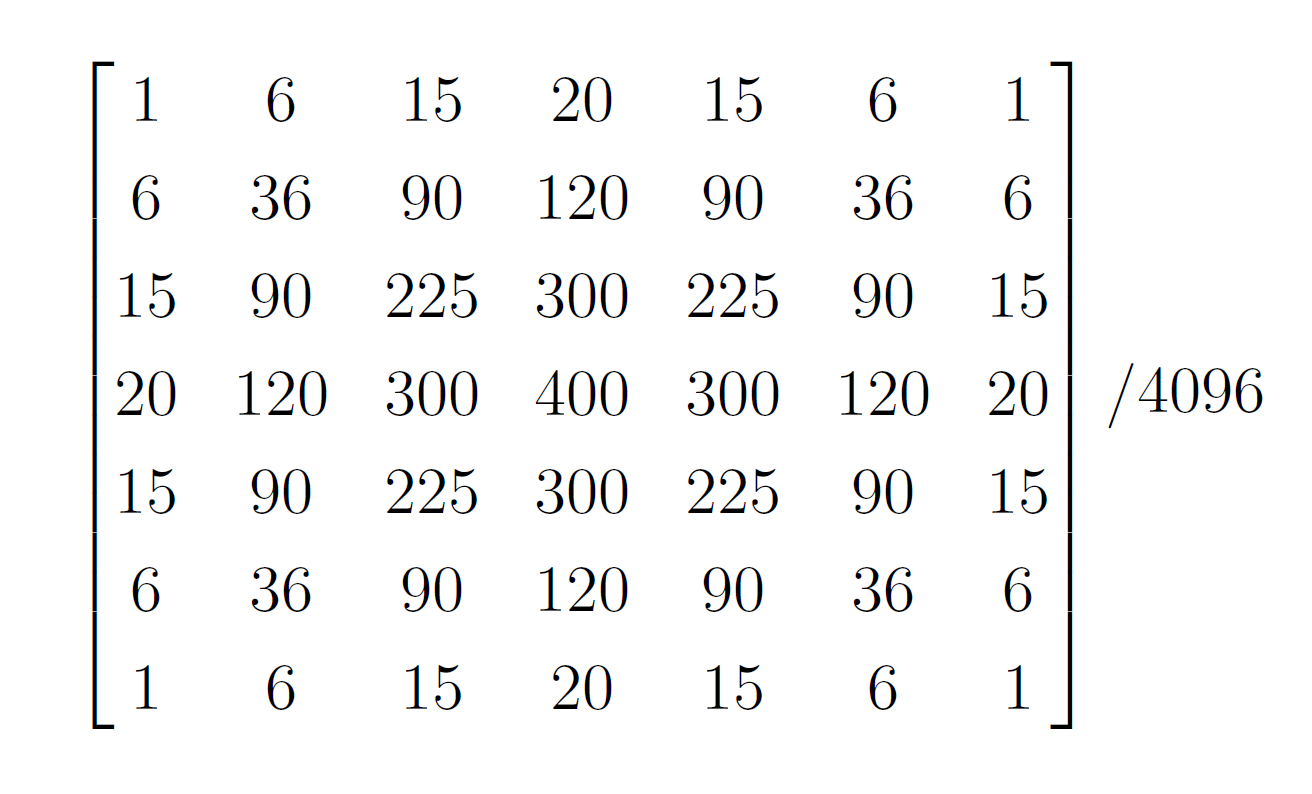}
     \caption{Kernel used for the Gaussian filter}
     \label{fig:gaussian_filter}
    \end{figure}

    Gaussian filtering is a convolution operation of the Gaussian kernel over the entire image. The original ORB-SLAM filter uses a filter size of 7x7 with standard deviation $\sigma = 2$. Upon experimentation, with various $\sigma$ values, there wasn't a noticeable change in accuracy. A Gaussian kernel with $\sigma = 2$ would involve fixed point multiplication, and require many DSPs. To make our filter more hardware friendly, we use a binomial Gaussian kernel of size 7x7 shown in Figure \ref{fig:gaussian_filter}. This filter is synthesized without any DSPs, because the multiplications have a constant multiplicand.
    
    \subsection {Orientation}
    The orientation requires a 37-row linebuffer as the input. It involves two parts. Calculating the moments $m_{10}$ and $m_{01}$ and using them to compute the arc tangent of $\frac{m_{10}}{m_{01}}$.
    \subsubsection {Moment} The naive approach to calculate the moment would involve calculating the individual parts in equation \ref{eq:moment} every time. This would require storing a window buffer of 37x37 and either multiple cycles or multiple hardware resources in order to compute the moments. We implement a recursive approach to calculate the moments, where each moment can be computed from the last moment, the incoming column and the outgoing column. 
    
     For the following section, we define these terms:
     
     \begin{itemize}
         \item $m_{00}$, from equation \ref{eq:moment} as the sum of all the pixels in the 37x37 window.
         \item $C_{in}$, the input column into the module as a 37x1 vector.
         \item $C_{out}$, the outgoing column leaving the window. It is the same as $C_{in}$ delayed by 37 clock cycles. It is a 37x1 vector.
         \item $m_{10}(X)$,  $m_{01}(X)$ as functions that compute the contribution of a single column $X$ to either the x-moment or y-moment in equation 
         \item $S(X)$, the sum of the pixels in column $X$.
         \item $I(x,y)$, the intensity of the pixel at $(x,y)$.
     \end{itemize}
     
     \subsubsection{Y moment} The Y moment is straightforward to compute, as our window buffer slides along the x-dimension, leaving the y-dimension  unchanged. The Y moment of the window can be recursively defined as:
     
     $$ m_{01_{n+1}} = m_{01}(C_{in}) + m_{01_n} - m_{01}(C_{out}) $$
     $$ m_{01_0} = 0$$
     
     \subsubsection{X moment} The X-moment can be broken down by looking at how each incoming column $C_{in}$ contributes to the moment, and how it changes when the window  shifts by 1 pixel. All pixels in a column share the same x-coordinate, which means the sum of the moments of the column, is the same as the moment of  the sum of the column as shown in the equations below.
     
     The definition of the X-moment is:
     $$ m_{10} =  \sum_{x,y=-18}^{18} xI(x,y)$$
     The incoming column $C_{in}$ is at the left edge of the 37x37 window, which gives it the x-coordinate -18.
     $$ m_{10}(C_{in}) =  \sum_{y=-18}^{18} -18I(-18,y) = -18 \sum_{y=-18}^{18} I(-18,y) = -18S(C_{in})$$
     Similarly, the outgoing column $C_{out}$ has the x-coordinate as 18.
     $$ m_{10}(C_{out}) =  18 \sum_{y=-18}^{18} I(18,y) = 18S(C_{out})$$
     
     Every time the pixels shift by 1, the x-coordinate of all the pixels in the 37x37 window increases by one. Effectively, the moment contribution of each column increases by the sum of the column, i.e $m_{10_{n+1}}(C_n) = m_{10_n}(C_n) + S(C_n)$. This contribution can be simplified across the entire window by keeping a running sum of all the pixels in the window defined as $m_{00}$.
     $$ m_{00_{n+1}} = m_{00_n} + S(C_{in}) - S(C_{out}) $$ where
     $$ m_{00_0} = 0 $$
     
     Using the above definition, we can recursively define the X-moment as follows:
     $$ m_{10_{n+1}} = m_{10_n} - 18S(C_{in}) - 18S(C_{out}) + (m_{00_n} - S(C_{out}))$$
     where
     $$m_{10_0} = 0$$
     which can be further simplified
     $$ m_{10_{n+1}} = m_{10_n} - 18S(C_{in}) - 18S(C_{out}) + m_{00_n} - S(C_{out}) $$
     $$ m_{10_{n+1}} = m_{10_n} -18S(C_{in}) - 19S(C_{out}) + m_{00_n}$$
    
     Pipelined adder trees are used to compute $S(C_{in})$ as it is shifted in. To save on having to recompute the $S(C_{out})$, we use a shift register to delay $S(C_{in})$ by 37 clock cycles.
     
    \subsubsection{Angle}
    \label{section:angle}
\begin{figure}[h]
        \centering
            \begin{subfigure}[b]{0.4\linewidth}
            \centering
            \includegraphics[width=\linewidth]{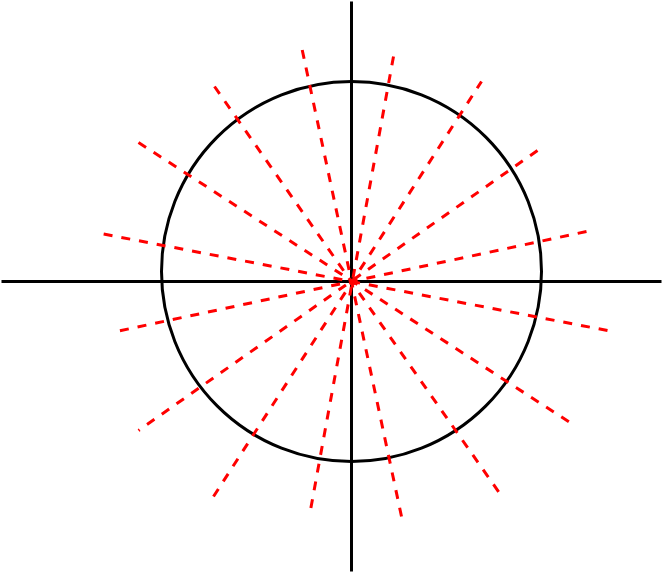}
            \subcaption{Example of angle space being discretized into 16 sectors}
            \label{fig:sectors}
        \end{subfigure}
        \hfill
        \begin{subfigure}[b]{0.4\linewidth}
          \centering
          \includegraphics[width=\linewidth]{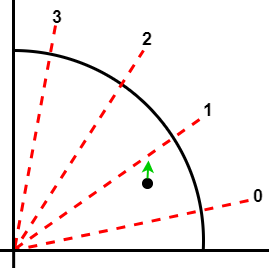}
          \subcaption{4 sectors per quadrant. Black dot, representing a keypoint is rounded to the nearest sector, 1}
          \label{fig:quadrant}
        \end{subfigure}
         \hfill
         \caption {Discretization of angle space into sectors}
     \end{figure}

    Using the moments calculated from the previous block, we compute $\theta = arctan(m_{01}/m_{10})$ as the angle of the intensity centroid. The computation involves division and arc tangent, which are time-consuming operations in hardware. There exist CORDIC engines that take several cycles and multiple DSPs to compute arctangents with relatively precise outputs. However, CORDIC engines are restrictive in the size of the input $x$ and $y$, which $m_{01}$ and $m_{10}$ generally exceed. Lookup Table (LUT) based methods can take 10s of cycles while using multiple BRAMs and DSPs to output an angle with a precision of less than 0.2 degrees. The feature extraction algorithm does not need to be precise, and the amount of resources required can be drastically reduced by discretizing the angles that a keypoint can have. We divide the circle into $N_D$ sectors and round up the angle to the nearest sector. The sectors are defined by the line that divides the sectors in half. Figure \ref{fig:sectors} shows an example of the circle being divided into 16 sectors. 
    The intensity centroid of a keypoint would be rounded up to lie on one of the red dotted lines on the circle. 
    
\begin{figure}[h]
        \centering
        \includegraphics[width=\linewidth]{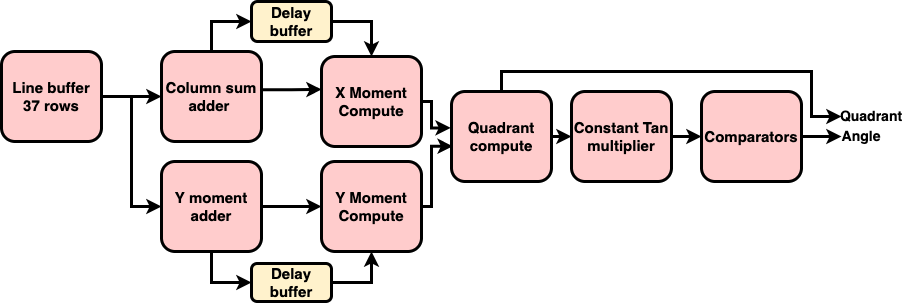}
        \caption{Block diagram of Orientation module}
        \label{fig:orientaion_block_diagarm}
    \end{figure}

    We determine which quadrant the centroid belongs to based on the sign of $m_{01}$ and $m_{10}$, and then shift the coordinates to the first quadrant by taking the absolute value of $m_{01}$ and $m_{10}$. As an example, if we discretize our angle space to 16 sectors, each quadrant will have 4 sectors with angles at 11.25, 33.75, 56.25, and 78.75 degrees as shown in Figure \ref{fig:quadrant}
    . To compute $\theta$, we approximate $\theta = arctan(y/x) = arctan(m_{01}/m_{10})$ using LUTs and comparators to approximately satisfy the following equation.
    $$ abs(m_{10}) * tan(\theta) = abs(m_{01}) $$
    We compute $ abs(m_{10}) * tan(\theta) $ by hardcoding the fixed point representations of all the sectors in the quadrant as shown in Table \ref{tab:tan_approximation}.
    Four comparators in parallel check if $abs(m_{10}) * tan(\theta) > abs(m_{01})$ and use a priority encoder to determine the closest match for $\theta$. Once determined, the module outputs the quadrant as well as $\theta$ to the BRIEF module. The architecture is illustrated in Figure \ref{fig:orientaion_block_diagarm}.
    
     \begin{table}[h]
         \caption{Tan approximation example, 4 sectors per quadrant}
         \label{tab:tan_approximation}
         \centering
         \begin{tabular}{|c|c|c|c|}
             \toprule
             $\theta$ (degrees) & $tan(\theta)$ & \specialcell{Fixed point\\ approximation} & \specialcell{Fixed point\\ representation} \\
             \midrule
              11.25 & 0.199 & $0.1875x$ & $(2^{-4}+2^{-3})x$\\
              33.75 & 0.668 & $0.65625x$ & $(2^{-5}+2^{-3}+2^{-1})x$\\
              56.25 & 1.496 & $1.5x$ & $(2^{-1}+2^0)x$ \\
              78.75 & 5.027 & $5x$ & $(2^{2} + 2^{1})x$\\
              \bottomrule
         \end{tabular}
     \end{table}
    
\subsection{rBRIEF}
To compute BRIEF descriptors, a 37x37 window is constantly updated using the input from the linebuffer. The orientation module outputs the angle and the coordinates of the point it processed. The coordinates from the orientation module is constantly compared with the head of the keypoint FIFO, and, if there is a match, the BRIEF module begins processing. While the keypoint is being processed, the 37x37 window is frozen until the descriptor is computed. Since BRIEF descriptors take multiple cycles to compute, if another keypoint is detected while the current ORB is processing, the keypoint is dropped unless the entire pipeline is stalled. In order to avoid dropping too many keypoints and to avoid stalling the pipeline to maintain a 1 pixel/cycle throughput, we instantiate multiple BRIEF modules that are controlled by an arbiter. BRIEF computations are dispatched to computation modules that are not busy. If all the modules are busy, and a new keypoint is detected, the keypoint is dropped. This is not a major issue, as keypoints are detected on average every 100-150 pixels and the SLAM algorithms are robust enough to handle a few keypoints being dropped. After the descriptor is generated, the module needs 37 cycles to reload the window, before it is ready to process a new keypoint.

The BRIEF module consists of a rotator and a generator module. The rotator computes the rotated coordinates of the BRIEF pattern at 1 pattern/cycle using equation \ref{eq:brief_rotation}. The generator then takes 1 cycle to lookup the pixels from the window buffer and 1 more cycle to generate the descriptor. The rotator requires the values of $cos(\theta)$ and $sin(\theta)$ to be passed in by the dispatcher. As the angles are discretized, we can use small lookup tables to store the cosine and sine values for all the possible angles in a quadrant. The $sin$ and $cos$ values are stored in 8-bit fixed point. Based on the quadrant, the sign of the outputs is adjusted. 

\begin{figure}[h]
    \centering
    \includegraphics[width=0.9\linewidth]{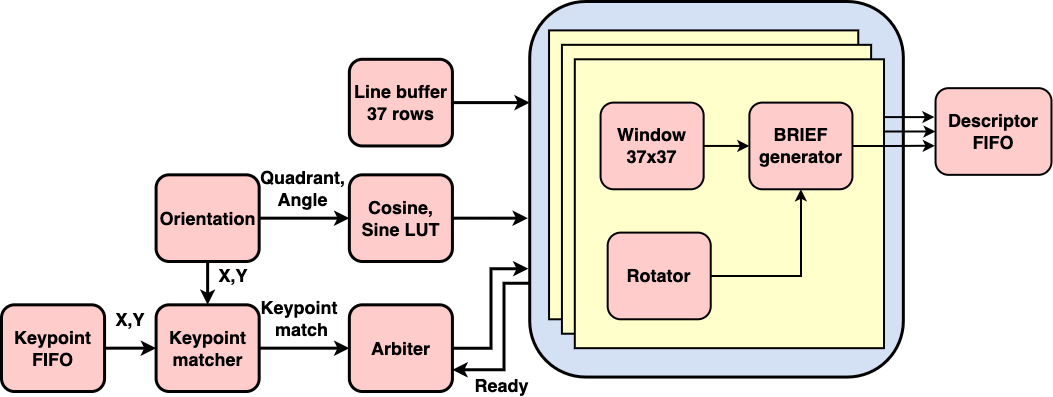}
    \caption{Block diagram of BRIEF module}
    \label{fig:brief_block_diagram}
\end{figure}

To eliminate the need for computing the rotated patterns, the original ORB paper proposes pre-computing the BRIEF patterns for all the possible angles and using a lookup-table during runtime to fetch the rotated coordinates. Each LUT would consume 256 pairs * 4 coordinates per pair * 5 bits per coordinate = 5120bits. If the angle is discretized into 32 sectors, the LUTs would requires 160Kbits of memory.  The computation cost of 3 cycles and a few DSPs  is insignificant compared to the resource usage of the LUT method, and hence we did not use the LUT method.

    \subsection {Feature matching}
    Similar to the works of Liu, et al. \cite{liu2019eslam}, we implement a heap that sorts the keypoints extracted by Harris scores, evicting the older keypoints back into DRAM. The feature matcher gets descriptors of the keypoints from the heap and computes the Hamming distance between the given keypoint and the keypoints mapped in the frame. If the computed Hamming distance is above the given threshold, the keypoint is "matched" and stored into a resulting buffer that is sent back to the CPU at the end of the frame.
    
    \subsection{Comparison of various hardware architectures}
    \label{section:other_works}
    Weberruss, et al. \cite{weberruss2017fpga} and Tran, et al. \cite{tran2018stream} both use Harris Corner detector instead of FAST as the former is more accurate but computationally more expensive. Through experimentation we determined that for SLAM applications, high success rate of corner detection is not critical to the accuracy.
    The BRIEF modules we implemented are similar to that implemented in Weberuss, et al.\cite{weberruss2017fpga}. The orientation computation is part of the BRIEF module, whereas we have the orientation module separated, avoiding duplication and saving on hardware resources. Similarly, we do not duplicate the cosine and sine LUTs, and instead pass the values in during dispatch.
    Fang, et al. \cite{fang2017fpga} use frame buffers to store the image in 2 levels. We use a streaming method, so no frame buffers are required, saving a considerable amount of BRAM.
    Liu, et al. \cite{liu2019eslam} utilize the same hardware for multiple scale levels. They have a dedicated image resizing module reading/writing directly from DRAM. This approach does not exploit the inherent parallelism available in processing multiple levels. It also incurs additional DRAM bandwidth for image resizing which we can avoid by using a streaming approach.
    
    \section {Experimental results and observations}
    \label{section:experiments}
    \textbf{\textit{Hardware setup}}: 
    The accelerator system is implemented on an Avnet Ultra96 development board. The system contains an Xilinx ZU3EG FPGA-SoC on-board with a quad-core ARM clocked at 1.5GHz, with the clock frequency of programmable logic running at 150MHz.

    \textbf{\textit{Dataset}}: We evaluate our SLAM system using the TUM dataset \cite{sturm12iros} which provides RGB camera data, along with depth information, obtained from a handheld Kinect moving through various office environments. It is widely used in the visual SLAM community to evaluate the accuracy of algorithms. The datasets provide the ground truth of the pose of the camera obtained using a high-accuracy motion-capture system. The images are provided at a resolution of 640x480 at 30fps. 
    
    We use Absolute Trajectory Error (ATE) as the metric to evaluate accuracy. It measures the absolute difference between the ground truth poses and the estimated poses, and outputs the mean, median, and standard deviation of these differences. The root mean squared error (RMSE) of these differences is well suited to evaluating the accuracy of visual SLAM systems. The ORB-SLAM algorithm is not deterministic and has a variance of about 10\% in the reported ATE across multiple runs using the same data. 
    
    We evaluate the accuracy and the latency of feature extraction of our implementation on various datasets and compare it with eSLAM \cite{liu2019eslam}. The latency of feature extraction in our implementation is around 2.5ms, which is 3.7x faster than eSLAM\cite{liu2019eslam}.  We also compare the resource utilization of our design as show in Table \ref{tab:resource_utilization}, where we extracted the resource utilization of eSLAM from the paper \cite{liu2019eslam}.
    
    \begin{table}[h]
        \caption{FPGA resource utilization of ORB feature extraction}
        \label{tab:resource_utilization}
        \centering
        \begin{tabular}{|c|c|c|c|c|}
        \toprule
         Implementation & LUTs & FFs & DSP  & BRAM \\
         \midrule
        eSLAM\cite{liu2019eslam} & \specialcell{56594} & \specialcell{67809} & \specialcell{111} & \specialcell{78}\\    
        This work &  76424 & 101694 & 80 & 120\\
        \bottomrule
        \end{tabular}
    \end{table}
    
    \subsection{Determining sectors for orientiation approximation}
            \label{section:sector_approximation}

\begin{figure}[h]
        \centering
        \includegraphics[width=\linewidth]{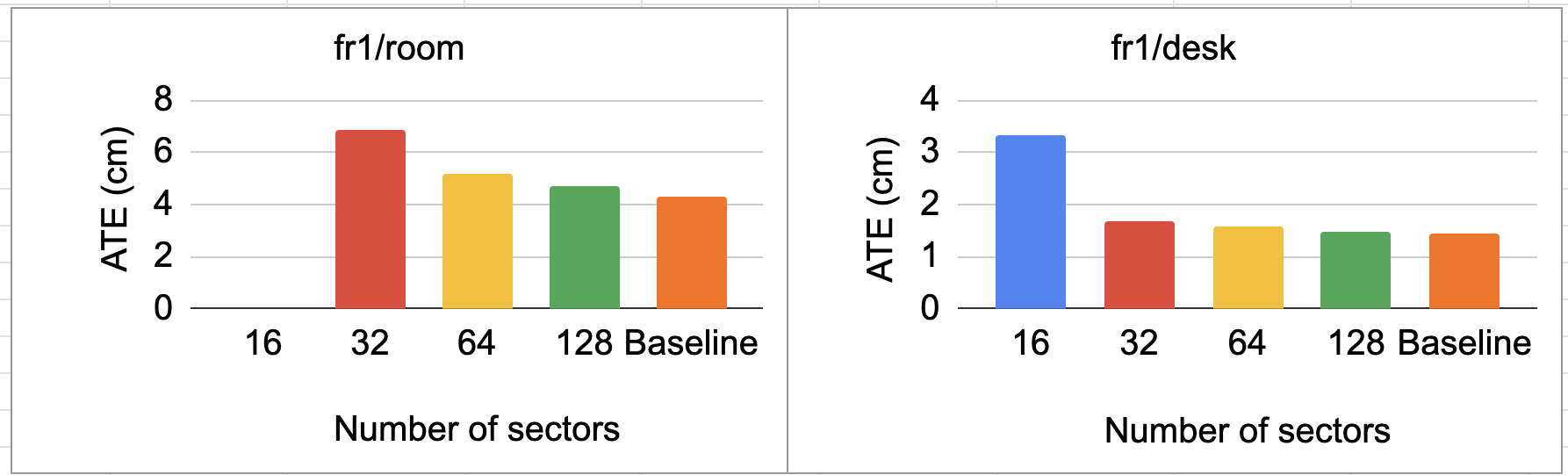}
        \caption{Effect on accuracy when discretizing the angles to various sectors across 2 datasets fr1/desk and fr1/room}
        \label{fig:sector_discretize_graph}
    \end{figure}

    The original ORB-SLAM implementation uses the opencv \textit{fastatan2} function to compute the orientation. The function has a precision upto 0.3 degrees. In our hardware implementation, we discretized the angles into sectors as described in Section \ref{section:angle}. In order to determine the appropriate amount of sectors to discretize the angle space, we collect data on the effects of accuracy vs number of sectors. As seen in Figure \ref{fig:sector_discretize_graph} we observe a trend where the accuracy greatly reduces between 64 sectors to 32 sectors. Based on this data, we discretized our orientations to 64 sectors.
    
    \subsection {Effect of input pixel bit-depth on accuracy}
        \label{section:pixel_depth}

    A large amount of Flip-Flop resources are consumed by the window buffers in the FPGA. A single ORB level uses approximately 38k FFs. In order to fit the design into the device, we explore reducing the bit-depth of each pixel at different stages of the ORB pipeline. We experimented with the pixels coming in to the FPGA from DRAM by truncating the lower bits after the FAST corner detection algorithm. We chose 2 datasets \textit{fr1/room} and \textit{fr1/desk} as they involve rapid movements, larger maps and loop closing to evaluate the impact on accuracy. The incoming pixels to the FPGA are 8-bit grayscale pixels.
    
    \begin{figure}[h]
        \centering
        \includegraphics[width=\linewidth]{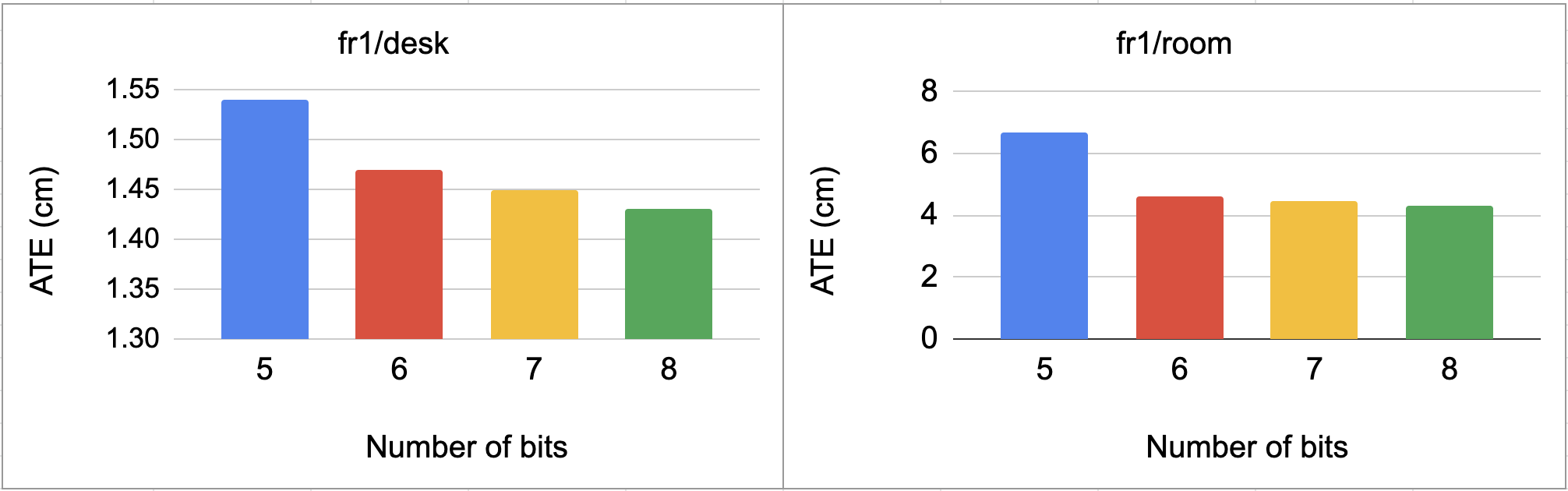}
        \caption{Effect on accuracy with varying pixel bit-depth}
        \label{fig:pixel_quant}
    \end{figure}
    
    Figure \ref{fig:pixel_quant} illustrates the loss in accuracy as the bit-depth reduces. We chose to quantize our pixels to 6 bits per pixel which offers us a 25\% reduction of FF resources while only dropping ~3.9\% in accuracy. However, this does not reduce the resource consumption of the BRAMs, as the way they are allocated is either as a 2Kx9bit array or a 4Kx4 bit array. The BRAM consumption is not a significant resource constraint, so this is not an issue.
    
    \subsection{Accuracy evaluation}
    \label{section:accuracy_evaluation}
\begin{figure}[h]
        \centering
        \includegraphics[width=\linewidth]{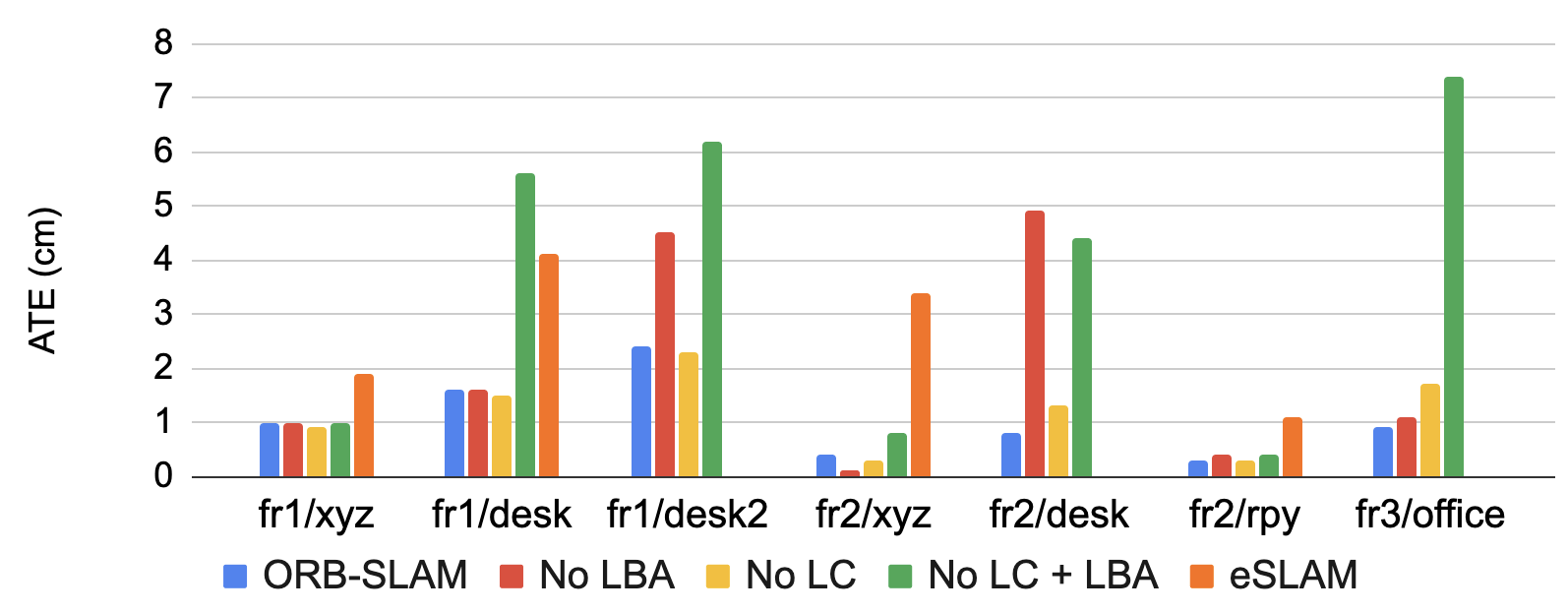}
        \caption{Tracking accuracy of datasets with Local Bundle Adjustment (LBA) and Loop Closing (LC)}
        \label{fig:tracking_accuracy}
    \end{figure}

    We compare our work mainly with eSLAM \cite{liu2019eslam} as it also implements an end-to-end ORB-based SLAM system on an SoC. We also compare our work against the original ORB-SLAM which was designed to run on a CPU.
    The runtime of our tracking thread on software is ~30ms, while eSLAM's is around 20ms. Based on an initial analysis, our tracking thread is far more complex than that of eSLAM. The exact algorithm is not specified in the paper, but it appears that they only perform motion-only BA without using local BA or loop closing. To further investigate the cause for the discrepancy in accuracy between eSLAM and CPU-based ORB-SLAM, we run ORB-SLAM with local BA and loop closing disabled on the various datasets. Figure \ref{fig:tracking_accuracy} plots the normalized accuracy for each dataset against the original ORB-SLAM, while Table \ref{tab:slam_accuracy} contains the raw data. We observe that without local BA, there is a large drop in accuracy in the datasets \textit{fr2/desk} and \textit{fr3/office}. Both of these involve extensive translatory motion and generate a large enough map such that keyframes are spread out enough where local BA makes a difference. 
    
    Loop closure also has an impact on those datasets where loops exist. In our case, \textit{fr1/room} and \textit{fr3/office}, both contain large loops. As seen in the chart, these two datasets have a significant accuracy drop, while the other datasets remain the same. Looking at the accuracy of \textit{fr1/room} in particular, the accuracy significantly drops when removing local bundle adjustment and loop closing. This dataset has the camera moving across a large open room, eventually ending back at the starting point. 
    
    eSLAM also proposed a novel rotationally-symmetrical BRIEF descriptor (RS-BRIEF) which reduces the hardware complexity in rotating the BRIEF descriptors by changing it from a trignometric operation to a bit rotation operation. This approach saves computation time and resources; however, it also leads to great loss in accuracy. The BRIEF patterns chosen in the ORB-SLAM algorithm were deliberate to maximize the accuracy as described in Section \ref{section:BRIEF}. To implement RS-BRIEF in our work, we would have to retrain the vocabulary used to describe the BRIEF descriptor. The amount of resources saved by using RS-BRIEF does not warrant the great drop in accuracy.
    
    \begin{table}[h]
        \caption{ATE (in cm) of various modifications to ORB-SLAM Tracking. (X indicates data was not available)}
        \label{tab:slam_accuracy}
        \centering
        \begin{tabular}{|c|c|c|c|c|c|}
        \toprule
         Dataset & ORB-SLAM & No LBA & No LC  & \specialcell{No \\LC or LBA} &eSLAM \\
         \midrule
        fr1/xyz & 1.0  & 1.0  & 0.9 & 1.0  & 1.9 \\
        fr1/desk& 1.6 & 1.6 & 1.5 & 5.6 & 4.1 \\
        fr1/room& 4.3 & 7.3 & 10.6 & 20.2 & 11.1\\ 
        fr1/desk2  & 2.4 & 4.5 & 2.3 & 6.2 & X \\
        fr2/xyz & 0.4 & 0.1 & 0.3 & 0.8 & 3.4 \\
        fr2/desk & 0.8 & 4.9 & 1.3 & 4.4 & X \\
        fr2/rpy  & 0.3 & 0.4 & 0.3 & 0.4 & 1.1 \\
        fr3/office & 0.9 & 1.08 & 1.7 & 7.4 & X\\
        \bottomrule
        \end{tabular}
    \end{table}
    
\begin{figure}[h]
        \centering
        \includegraphics[width=\linewidth]{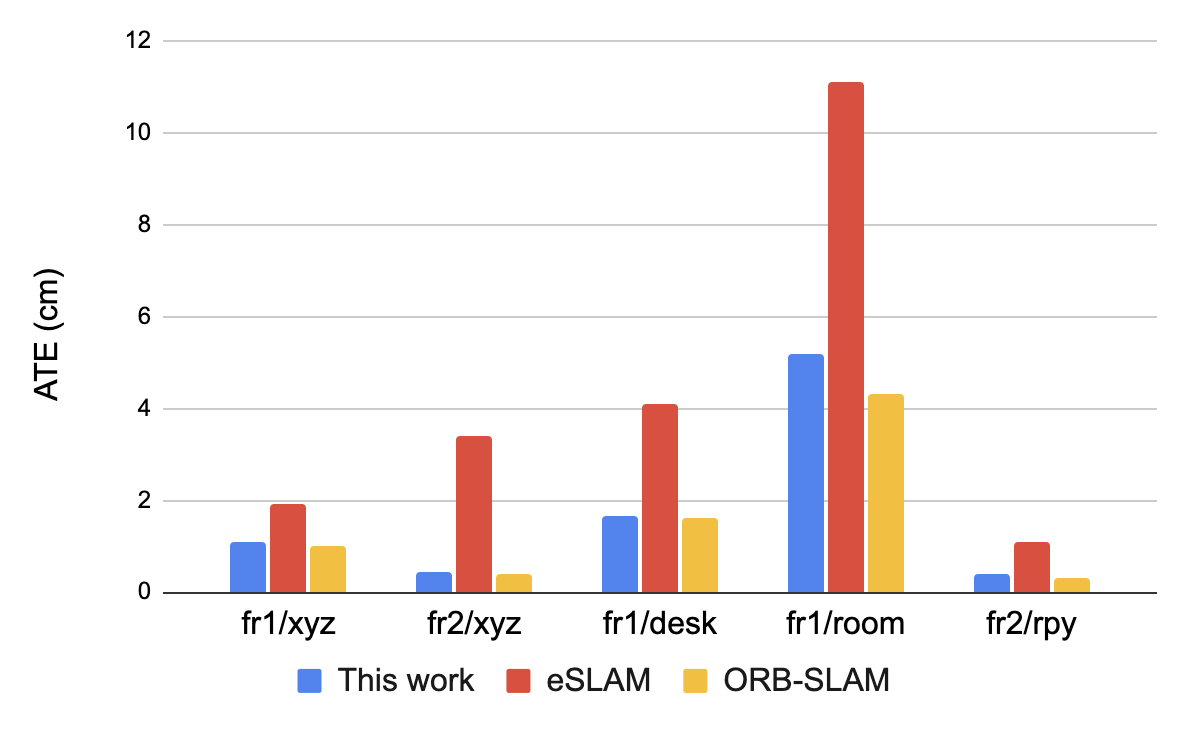}
        \caption{Accuracy comparison amongst different platforms on various datasets}
        \label{fig:accuracy comparison}
    \end{figure}

    Figure \ref{fig:accuracy comparison} illustrates the accuracy comparison of our work against eSLAM and the original CPU implementation of ORB-SLAM. Our work modified ORB-SLAM with the hardware optimizations described in Sections \ref{section:sector_approximation} and \ref{section:pixel_depth}. On average, our work is around 5-10\% less accurate compared to the CPU version of ORB-SLAM and 50-70\% more accurate when compared to eSLAM. This improvement in accuracy compared to eSLAM is due to the addition of Local BA and Loop closing threads which greatly improves accuracy in complex datasets. The degradation in accuracy compared to the original ORB-SLAM is acceptable as it is a relative comparison to the error rate. The absolute difference in observational error (averages around 0.01-1cm based on the dataset) is insignificant when compared to the total distance travelled (1.5m-20m).
    
    \subsection {Performance and Power Evaluation}
    
    The performance of this work is compared with software implementations on the integrated ARM processor of the SoC, a desktop with an Intel i5 processor coupled with a NVIDIA 2070S, and a Jetson Xavier evaluation platform that runs an embedded SoC. The Jetson SoC contains an 8-core ARM CPU and a 512-core GPU. The ARM CPU mentioned in the table runs the entire algorithm on the SoC's CPU. We also compare our work against the state-of-the-art eSLAM \cite{liu2019eslam}. Table \ref{tab:framerate_comparison} shows the runtime comparison of our work against the different evaluation platforms. The performance is a bit worse than the desktop GPU + CPU variant, however our system consumes orders of magnitude less power allowing it to be used in mobile environments. We are able to achieve a 1.55x speedup in comparison to the desktop CPU, and a 1.35x speedup compared to eSLAM. The table also shows the power consumed by the various platforms. Our work is 43x, 9.7x more power efficient when compared to desktop systems with and without a GPU respectively. Compared to the embedded platforms, our work is 2.1x more power efficient than the Jetson, while consuming 2.5x more power than the ARM CPU standalone. We consume 2.4x more power than eSLAM as the higher throughput and accuracy are reflected in the resource consumption, and computational complexity.
    
    \begin{table}[h]
        \caption{Framerate and power consumption of ORB-SLAM on various platforms}
        \label{tab:framerate_comparison}
        \centering
        \begin{tabular}{|c|c|c|c|}
        \toprule
         Platform & Framerate & Speedup & Power\\
         \midrule
        Desktop CPU & 40 & 1x & 45W\\
        Desktop GPU & 70 & 1.75x & 200W\\
        Jetson      & 22 & 0.55x  & 10W\\
        ARM CPU     & 7 &  0.175x & 1.8W\\ 
        eSLAM\cite{liu2019eslam} & 45 & 1.125x & 1.9W\\
        This work & 62 & 1.55x & 4.6W\\
        \bottomrule
        \end{tabular}
    \end{table}

    \section {Conclusion}
    \label{section:conclusion}
    In this paper, we presented an end-to-end implementation of ORB-SLAM on an FPGA to ensure both real-time performance and high accuracy. In order to meet performance requirements, the ORB accelerator is formatted as a streaming processor, which avoids the utilization of memories, enables data re-use and processes 1 pixel per cycle. To facilitate fitting the high-performance accelerator into an embedded device, data-driven hardware optimizations were made by trading-off hardware resources for lower accuracy. The optimizations were made based on experimental observations to maximize resource savings, while minimizing accuracy loss. Leveraging the above factors, the evaluation results show that we achieve a 1.55x speedup compared to a desktop CPU while maintaining a reasonable accuracy, and 1.35x speedup compared to previous works \cite{liu2019eslam} while averaging a 2x improvement in accuracy across multiple datasets.

\bibliographystyle{./bibliography/ACM-Reference-Format}
\bibliography{./bibliography/ref}

\end{document}